# On the Study of Collective Dynamics in Supercooled Liquids through the Statistics of the Iso-Configurational Ensemble

Asaph Widmer-Cooper and Peter Harrowell

School of Chemistry, University of Sydney, Sydney,
New South Wales, 2006, Australia

**Abstract**

The use of the iso-configurational ensemble to explore structure-dynamic correlations in supercooled liquids is examined. The statistical error of the dynamic propensity and its spatial distribution are determined. We present the spatial distribution of the particle non-Gaussian parameter as a measure of the intermittency with which particles exhibit their propensity for motion. The ensemble average of the direction of particle motion is introduced to establish the anisotropy of the dynamic propensity.

# 1. Introduction

Collective dynamics can be characterised by the spatial distribution of particle movement. This observation is old news in the case of harmonic solids where the particle momenta exhibit significant spatial correlations in the form of normal modes. In the case of amorphous solids, these collective movements can be complex indeed[1]. What is new is the idea that diffusive motion, occurring over time scales three or more orders of magnitude longer than that required for the relaxation of momenta correlations, can also exhibit interesting spatial distributions. This is the case, generally, in supercooled liquids[2-5] and in some structured liquids at or above their freezing temperatures[6]. The explicit spatial information about the collective motions in these disordered systems encoded in the spatially distributed kinetics promises to go some way to dispelling the fog that has settled around words like "caging", "free volume" and "hopping", words which remain common descriptions of collective behavior in liquids.

In the case of normal modes, the connection between the dynamical structure and the particle configuration is explicit, with the link provided by the force constant matrix. In the case of a harmonic solid, therefore, we do not need to actually carry out a dynamic calculation to describe the collective motion. The situation is quite different for liquids that exhibit persistent dynamic heterogeneities. Such heterogeneities typically involve large amplitude motion by a subset of particles that preclude a harmonic description. There is, to date, no equivalent of the normal mode analysis with which to predict collective large amplitude motions from a given arrangement of particles. The spatial information provided by the dynamics is therefore the sole source of information about collective behaviour at this time, and so represents a fundamental object of interest to those wanting to understand the relationship between structure and dynamics in disordered materials. Specifically, we are interested in establishing the casual link between the particle configuration and the spatial distribution of dynamics to which it gives rise. Only here we must deal with the inverse problem: given a particular distribution of dynamics, what can we learn about the particle arrangements that are responsible?

How do we establish what information is significant and what is noise? In this paper, we describe a method of removing the influence of the initial momenta distribution from the resulting map of mobilities through the introduction of a property of a configuration called the "propensity for motion". The propensity was introduced in a 2004 letter[7] and this paper represents the first extended discussion of the method. The statistical treatment of the large fluctuations in particle mobilities in supercooled liquids is an important and non-trivial problem. In this paper we shall also discuss the statistical significance of the spatial distribution of the dynamic propensity.

The paper is organised as follows. Section 2 describes the propensity and the specific model system used in this work. In Section 3 we investigate the effect of varying configuration and temperature, and in Section 4 we consider the statistical significance of the propensity maps. Section 5 then examines the variability of single particle motion and the intermittent manner in which configurational influence is expressed. In Section 6 we introduce a number of correlations within the iso-configurational ensemble and the information they can provide about the character of collective motion. Our conclusions are presented in the final section.

**2. Defining the Dynamic Propensity of a Particle**

Consider, first, the possibility that there is no correlation at all between an initial configuration and the subsequent particle dynamics. In this case, each particle's squared displacement, averaged over many trajectories with the same initial configuration, would be the same as that of every other particle of the same species. The trajectory averaging is necessary to remove the uninteresting variation in the particle displacements arising from the choice of initial momenta. Conversely, if we find that there *are* significant differences in the trajectory-averaged squared displacements of different particles, then we have identified a feature of the dynamics that can be directly attributed to some aspect of the initial configuration.

To this end, we have introduced the iso-configurational ensemble consisting of $N_{runs}$ separate simulation runs over a fixed time interval, all starting from the same particle configuration but with momenta randomly assigned from the Maxwell-Boltzmann distribution at the appropriate temperature. Let $f_i(\Delta r)$ be the ensemble distribution of the displacement of particle $i$ over the fixed time interval. These distributions represent the ensemble characterisation of each particle's capacity for movement from a specific initial configuration. They are, by construction, invariant to time reversal. We refer to the second moment of $f_i(\Delta r)$, i.e. the ensemble average of the squared displacement of particle $i$, $<\Delta r_i^2>_{ic}$, as the *dynamic propensity* of particle $i$ in the given initial configuration. The expression $\langle...\rangle_{ic}$ indicates an average over the iso-configurational ensemble. We shall consider other statistics of the distribution $f_i(\Delta r)$ in Sections 5 and 6.

This procedure does not constrain the total energy of the initial state and so we run these calculations in the NPT ensemble. The propensity, as defined here, can not be directly associated with the equilibrium distribution of trajectories passing through the initial configuration. This is because our assignment of the initial momenta neglects the correlations that exist between the instantaneous moment and the particle positions. (Despite this neglect, we note that the distribution of total energy among particles in the initial states with random momenta closely matches that obtained by the integration of the equations of motion, as shown in Figure 1 for the soft disc mixture.) We introduce the iso-configurational ensemble here as a tool to characterise a particle's tendency for motion, not as a rigorous method of calculating its equilibrium probability for motion.

To compare propensities from different temperatures $T$, we set the run time for a given trajectory to be 1.5 times the structural relaxation time $\tau_e$ ($\tau_e$ is defined in terms of the intermediate incoherent scattering function $F(k,t)$ such that $F(k_{max}, \tau_e)=1/e$, where $k_{max}$ is the wavevector of the Bragg peak and $e = 2.7183$, the base of the natural logarithm). This run interval was chosen to maximise the dynamic contrast between particles.

Since the 2004 paper,[7] there have been a number of applications of the idea of propensities to short time behaviour in glass forming systems. We have shown[8] that much of the spatial structure observed in a propensity map can be recovered by determining the spatial distribution of short time motion as measured by a Debye-Waller factor averaged over the iso-configurational ensemble. A related result has been reported by Appignanesi et al.[9] for a binary Lennard-Jones mixture in 3D. Instead of using the displacement $\Delta r_i = r_i(\tau_{run}) - r_i(0)$ to calculate the propensity as described above, these workers have used $\Delta r_i = r_i(\tau_{run}) - r_i(t^*)$, where $t^*$ is a variable delay time after the initial state. By recording how the spatial variation of propensity decayed as $t^*$ was increased, it was shown that most of the spatial correlations are the result of short time motion.

For a glass-forming liquid, we use a two-dimensional (2D) equimolar binary mixture of particles interacting via purely repulsive potentials of the form

$$u_{ab}(r) = \varepsilon \left[ \frac{\sigma_{ab}}{r} \right]^{12} \qquad (1)$$

where $\sigma_{12} = 1.2 \times \sigma_{11}$ and $\sigma_{22} = 1.4 \times \sigma_{11}$. All units quoted will be reduced so that $\sigma_{11} = \varepsilon = m = 1.0$ where m is the mass of both types of particle. Specifically, the reduced unit of time is given by $\tau = \sigma_{11} (m/\varepsilon)^{1/2}$. A total of $N = 1024$ particles were enclosed in a square box with periodic boundary conditions. This model and its approach to the glass transition have been studied in detail and readers are directed to these papers[10-13] for more details.

To visualise the spatial distribution of the propensity, it is useful to remove the additional complexity of the configuration and use contour plots for the propensity.

As the data points are located at irregularly spaced particle coordinates, it is necessary to interpolate between them. We found the modified version of Shepard's method[14] to be a useful algorithm for obtaining good fits to the data without introducing erroneous peaks and valleys. The occasional inconsistencies introduced by the interpolation near the periodic boundaries can be removed by fitting to a set of coordinates containing periodic images.

## 3. The Effect of Varying Configuration and Temperature

Even within a single temperature, there will be variation in the spatial structure and degree of heterogeneity from configuration to configuration. We generated ten configurations each at $T = 0.4, 0.46, 0.5, 0.55, 0.6, 0.8$ and the configurations were separated by $75\tau_e$ to ensure that they were significantly different from each other. The propensities for each initial configuration were averaged over 100 runs. The values of $\tau_e$ for the above temperatures are 673, 51.7, 13.6, 4.3, 2.9, 1.2 and 0.8, respectively.[10]

In Figure 2, we compare the mean, range (difference between max and min values), standard deviation (stdev) and the ratio stdev/mean for the propensity distribution calculated for each configuration. The most obvious change is a rapid increase in the range and standard deviation below $T = 0.5$, with the mean showing a smaller increase at low temperature. These changes are accompanied by an increase in the variation between different isothermal configurations. We conclude that below $T = 0.5$, there is a strong increase in the effect that the specific structure of a configuration has on the dynamics.

In Figure 3 we plot the propensity distributions separately for small and large particles for individual configurations at $T = 0.4$ and 1. At $T = 1$ the distributions are very narrow and quite similar for both particle species, but as the temperature decreases the distributions become broader and the small and large particle distributions increasingly differ. The distributions still overlap, but on average the small particles have higher propensity than the large ones.

A change in temperature or configuration also affects the spatial variation of propensity. The propensity maps for four configurations at $T = 0.4$ are shown in Figure 4. As expected, the distribution of high and low propensity regions varies significantly from plot to plot. There is an increase in the clustering of particles with similar mobility below $T = 1$. To better quantify this, we consider the aggregation of high propensity particles using a cluster analysis. For each configuration we select the 10% of particles with the highest propensities and assign them to clusters depending on whether they are a nearest neighbour to another particle already in a cluster. When all particles have been assigned to clusters, we count the total number of clusters and the variance in cluster size, and use these two quantities to characterise the degree of spatial clustering. Figure 5 shows the results of the cluster analysis for ten configurations each at $T = 0.4, 0.5, 0.6, 0.8$ and 1.

**4. The Statistical Significance of Propensity Structure**

The usefulness of the propensity as a measure of structure-related dynamics is due, directly, to the large variation in particle displacements that can be observed from run to run. Large variations, however, typically require large sample sizes if one wants high accuracy for the statistics of the distribution. In this section we investigate the uncertainty in the propensity distribution as a function of the number of runs and discuss the significance of the spatial variations depicted in the propensity maps.

The uncertainty in the propensity of particle $i$ is measured by the standard error $\sigma_i / \sqrt{N_{runs}}$, where $\sigma_i = \left( <\Delta r_i^4>_{ic} - <\Delta r_i^2>_{ic}^2 \right)^{1/2}$ is the standard deviation in the squared displacement distribution for particle $i$. The application of the Central Limit Theorem to highly asymmetric distributions such as those exhibited by the high propensity particles requires some care. For T = 0.4, we divide 1000 runs into independent non-overlapping sample sets, and obtain an approximately Gaussian distribution for sample sizes greater than 40, even for particles with the highest relative variance. We are therefore confident that the standard error provides a meaningful measure of uncertainty in the propensity for ensembles of 40 or more runs. We note that all our analysis has been performed on ensembles of at least 50 runs and usually of 100 runs.

To investigate the convergence of the individual propensities, we define the relative uncertainty in the propensity at the 95% confidence level and study its convergence as the number of runs increases. The $P\%$ confidence interval is defined as the interval in which there is a $P\%$ chance of finding the true population mean. To calculate the confidence interval for the propensity one should strictly use the two-sided Student's $t$-distribution[15] since the population mean and variance are unknown. However, in practice we find that the sample size, i.e. the number of runs, is sufficiently large that we can use the normal distribution instead. The relative uncertainty $R_i$ in the propensity of a particle *i* at the 95% confidence level as a function of the number of runs is, for a normal distribution, given by

$$R_i(N_{runs}) = 1.649 \frac{\sigma_i}{<\Delta r_i^2>_{ic} \sqrt{N_{runs}}} \qquad (2)$$

In Figure 6, we plot the mean uncertainty <*R*> (*R*$_i$ averaged over the *N* particles) as a function of the total number of runs for configurations at *T* = 0.4 and *T* = 1. The error bars indicate the range of *R*$_i$ values at a given *N*$_{runs}$. At *T* = 0.4, we find that while <*R*> has decreased to about 0.2 (i.e. 20%) after 200 runs, the maximum value decreases much slower, e.g. there are still some particles with *R*$_i$ = 0.6. By 1000 runs <*R*> has decreased to 0.1, but the largest relative uncertainties are still around 25%. In comparison, the uncertainty decreases much faster at *T* = 1. After 200 runs <*R*> = 0.12 and the maximum uncertainty is about 20%, and by 1000 runs <*R*> = 0.06 and the maximum is around 8%.

A thousand runs, each over a time interval of 1.5$\tau_e$, represents a formidable commitment for a single propensity map, particularly at low temperatures where $\tau_e$ is large. Fortunately, we are more interested in obtaining reliable coarse grained spatial maps and distributions of propensity magnitudes than being able to resolve variations between individual particles. We find, for example, that it takes only 100 runs, at both *T* = 0.4 and *T* = 1, for the standard deviation of the *total* propensity distribution to converge to within 2% of the extrapolated limit at infinite number of runs.

In Figure 7 we compare the spatial distribution of propensity averaged over ensembles of 50 and 1000 runs for the same configuration at $T = 0.4$. Although there are minor differences between the two plots, it is clear that the coarse grained spatial variation is established within 50 runs. The reason why the *spatial* distribution of propensity converges far more rapidly is that the difference between high and low propensities, i.e. the range of the propensity distribution, is generally much larger than the mean. In Figure 8 we plot a 1D traverse of a propensity map to provide some comparison between the standard error of each particle and the range of the distribution. As long as the latter is larger than the former, the coarse grained structure of the propensity map will be reliable. In fact, as shown in Figure 2a, the range increases rapidly relative to the mean below $T = 0.5$. If this rate is faster than the rate at which the uncertainty in the propensity increases, which it appears to be, then the spatial distribution of propensity should converge even more rapidly at lower temperatures. It is an attractive idea that this may make propensity calculations practical at deeper supercoolings than we have studied here.

## 5. Single Particle Variability

The variation in an individual particle's mobility between runs is important, not just because it influences the convergence properties of the propensity, but because it provides insight into the way the configuration influences the dynamics. A large variation in particle mobility between runs indicates a significant randomness in the manner in which the configuration influences the dynamics at low temperature. While there is a higher probability of a release event occurring in a high propensity region, both high and low propensity regions are capable of constraining the particles from moving in a given run. In other words, the configuration expresses its character intermittently.

The large variances of the individual particles are typically associated with highly asymmetric displacement distributions $f_i(\Delta r)$, with a peak at a low value of $\Delta r$ and a long tail extending to large displacements, as shown in Figure 9 for a representative particle at $T = 0.4$. This asymmetry can be quantified as a deviation from a Gaussian form through the use of a non-Gaussian parameter $\alpha_i$ for particle $i$ given by

$$\alpha_i = \frac{<\Delta r_i^4>_{ic}}{2<\Delta r_i^2>_{ic}^2} - 1 \qquad (3)$$

The quantity $\alpha_i$ equals zero for a Gaussian distribution. The $\alpha_i$ distributions for configurations at $T = 1.0$ and $0.4$ are plotted in Figure 10. While all the individual $f_i(\Delta r)$ distributions are close to Gaussian at high temperature, the supercooled sample exhibits a broad distribution of $\alpha_i$ values with most particles exhibiting a significantly non-Gaussian distribution of displacements. Note that this non-Gaussian parameter is quite distinct from that discussed previously in the context of supercooled liquids.[16,17] The $\alpha_i$ introduced here refers to the variety of displacements achieved by a *single* particle over the ensemble of trajectories, as opposed to the variety of displacements achieved by different particles in a single trajectory.

The non-Gaussian $f_i(\Delta r)$ distributions represent a new piece of kinetic information and the spatial distribution of single-particle non-Gaussian parameters $\alpha_i$, defined in Eq. 3, may offer additional insight into the manner in which the configuration influences relaxation. Those particles with motion that varies the most from run to run can be thought of as having the least structural constraint on their mobility. Figure 11 shows the spatial distribution of $\alpha_i$ and propensity for a configuration at $T = 0.4$. Regions with high propensity - and large particles with 6 large neighbours - tend to have low $\alpha_i$, and particles with high $\alpha_i$ tend to have low propensity. Lines of high $\alpha_i$ appear to represent paths for rare motion in regions of low propensity, and therefore a study of these may provide insight into mechanisms for relaxation of the slow regions. [If leave mention of large particles with 6 large neighbours here and in caption to figure 11 then need to use the figure with the large particles indicated]

Other physical phenomena in which rare events have a significant influence on some mean property have previously been described as `intermittent'. In particular, the term *intermittency* has been used to describe distributions in which maxima in space or time are widely spaced and rare, but make a dominating contribution to the physical quantity of interest. Ciliberto et al.[18] have reported intermittent voltage noise signals characterised by rare large noise spikes above the regular fluctuations during dielectric studies of a colloidal glass. The result is a non-Gaussian distribution for the voltage noise which, based on numerical work, has been interpreted in terms of activated and spontaneous relaxation events.[19] A review of recent experimental, numerical and theoretical work on the intermittency of relaxation in glassy soft matter can be found in Ref. 20.

There are two related types of intermittency at play in a supercooled liquid. In the course of a single trajectory, a subset of particles may exhibit high mobility for a period and then become immobile. Within the iso–configurational ensemble, a particular subset of particles may exhibit mobility in some runs but not others. The latter observation implies the former, but not vice versa. One way of interpreting the intermittent manner in which the configuration affects the dynamics is in terms of `constrain' and `release' events. Even high propensity regions are able to `constrain' particles, i.e. to not allow significant motion to occur, but occasionally `release' occurs, i.e. large displacements take place during a run. The difference between high and low propensity regions is that the frequency (or probability) of release is higher in the high propensity regions.

## 6. Correlations in the Iso-Configurational Ensemble

The iso-configurational ensemble allows for a variety of novel correlations. For example, the ensemble average of the dot product of a particle's displacement in different runs can give information on the degree to which the original configuration confers directionality on particle motion. We define the directionality $d_i$ of a particle $i$ as the mean dot product over all pairs of displacement vectors normalised by the propensity, i.e.

$$d_i = \frac{\frac{1}{N_{\alpha\beta}} \sum_\alpha \sum_{\beta \succ \alpha} (\Delta \vec{r}_{i,\alpha} \cdot \Delta \vec{r}_{i,\beta})}{<\Delta r_i^2>} \qquad (4)$$

where α and β are run indices, $N_{\alpha\beta} = {}^{N_{runs}}C_2$ is the number of distinct pairs of runs in the iso-configurational ensemble, $\Delta \vec{r}_{i,\beta}$ is the displacement vector of particle $i$ in run β and $<\Delta r_i^2>$ is the propensity of particle $i$. For a random distribution of displacements, the vector pairs will be evenly distributed in magnitude and $d_i = 0$. If the particle moves in the same direction in every run, $d_i \approx 1$.

In Figure 12 we plot $d_i$ against propensity, using data pooled from ten configurations each, at $T = 0.4$ and $T = 1$. The configurations were separated from one another by $75\tau_e$, and the propensities and directionalities were calculated over ensembles of 100 runs. At $T = 0.4$ the particles with high propensity generally have low directionality, suggesting that any directionality conferred by the initial configuration is rapidly `forgotten' as a particle moves away from its initial position. Surprisingly, we find relatively high directionality associated with low propensity particles. To understand these results we have looked at the distribution of displacement magnitudes and angles for individual particles. Two such maps are displayed in Figure 13 – one for a high directionality low propensity particle and the other for a high directionality high propensity particle. We conclude that much of the directionality measured here is arising from the particles that are initially displaced, by a small distance, from the local potential minimum. By the end of the run they are more likely to be near the minimum than not, hence the high directionality. This effect is most marked for low propensity particles because these have the most stable local minimum that retains their position from run to run. We conclude that the configuration imparts little directionality to the large amplitude displacements of the particles.[add comment here?]

We have employed the iso-configurational ensemble to clarify the connection between dynamics and structure. Such a connection also implies a correlation of the motion between particles via the particle configuration, i.e. a fluctuation in the structure resulting from a local set of particle displacements may well produce a high propensity domain that, in turn, results in further local displacements. The resulting dynamic correlations can be addressed within the iso-configurational ensemble. One can ask, for example, over what distance do such dynamical correlations persist? We have addressed this question as follows. Consider two particles, *i* and *j*, each with a distribution of displacements arising out of the iso-configurational ensemble. We would like to test for a correlation between the distribution of the magnitudes of the displacements. To do this we have calculated the Pearson's correlation coefficient $K^{15}$ over 100 runs where

$$K = \frac{\text{cov}(X,Y)}{\sigma_X \sigma_Y} \qquad (5)$$

and $X = |\vec{r}_i|, Y = |\vec{r}_j|$ are the displacement magnitudes of particles *i* and *j* and *cov(X,Y)* is the covariance of *X* and *Y*. Values of $K \approx 1$ imply a linear relation between the displacements of particles *i* and *j*, and if there is no linear correlation then $K \approx 0$. In Figure 14 we have plotted the values of *K* as function of the distance between particles *i* and *j* for a single configuration at *T* = 0.4. In this example, a single high propensity particle *i* was selected. We find a kinetic correlation length for the selected particle *i* of 2-3$\sigma_{11}$. This length is consistent with previous estimates of kinetic correlation lengths.[21]

Other types of ensemble averages are possible. To establish, for example, the kinetic significance of a particular group of mobile particles, we have compared the propensity for that subset of runs in which the selected particles showed small displacements with the propensities for all the runs.[22] This approach allows one to restrict the motion of a group of particles without perturbing the Hamiltonian. While gathering sufficient statistics can be a problem, this approach represents a quite general tool for posing questions about cause and effect in collective dynamics.

## 7. Summary and Conclusions

In this paper, we have explored the use of the iso-configurational ensemble in the study of collective dynamics in a supercooled liquid. The benefit of this approach is that it allows for the explicit expression of the relationship between the particular configuration that selects the ensemble and any statistic of the associated distribution of particle dynamics. In this paper we have introduced four such statistics: the particle propensity, the particle non-Gaussian parameter, the particle directionality and the radial pair displacement correlation function. Each statistic provides the means of articulating and answering a specific question concerning the spatial distribution of dynamics in a supercooled liquid.

In the case of the propensity, we answer the question, "what aspect of the particle mobility is determined by an initial configuration?" While the error in individual propensities can converge quite slowly in terms of the size of the sample of trajectories, we have shown that the error in the maps of propensity converge quite quickly, thanks to the large range of the propensities. The particle non-Gaussian parameter answers the question, "how intermittent is particle motion within the iso-configurational ensemble and what aspect of the particle intermittency is determined by an initial configuration?" We find that the spatial distribution of intermittency can differ quite markedly from that of the propensity for the same configuration. The question "to what degree does a configuration impart favoured directions of motion?" is answered with the ensemble averaged directionality. For the 2D glass-forming mixture studied here we find that the only directionality is that associated with relaxation of the initial particle displacements to their local energy minimum in the most stable of particle domains. Finally, in the radial pair displacement correlation function we have introduced an approach to characterising the time evolution of the ensemble displacement distributions through the local correlation between mobility (i.e. ensemble evolution) at one position and its dynamic consequences in the surrounding material. We hope that the description of these tools will help refine and inform the discussion of collective dynamics and encourage the development of yet other tools to address questions not yet considered.

**Acknowledgements**

We gratefully acknowledge helpful discussions with Toby Hudson. This work was supported by funding from the Australian Research Council.

**References**

1. J.P.Wittmer, A.Tanguy, J.-L. Barrat and L.Lewis, Europhys. Lett. 57, 423 (2002).
2. S.Butler and P.Harrowell, J.Chem.Phys. 95, 4466 (1991).

3. K.Schmidt-Rohr and H.W.Spiess, Phys.Rev.Lett. 66, 3020 (1991)

4. M.D.Ediger, Ann.Rev.Phys.Chem. 51, 99 (2000)

5. R.Richert, J.Phys.Cond.Matt. 24, R703 (2002).

6. M. Hurely and P. Harrowell, Phys.Rev.E 52, 1694 (1995).

7. A. Widmer-Cooper, P. Harrowell and H. Fynewever, Phys. Rev. Lett. 93, 135701 (2004)

8. A. Widmer-Cooper and P. Harrowell, Phys. Rev. Lett. 96, 185701 (2006)

9. G.A.Appignanesi, J.A.Rodriguez Fris, M.A. Frechero, Phys. Rev. Lett. 96, 237803 (2006).

10. D. Perera and P. Harrowell, J. Chem. Phys. 111, 5441 (1999)

11. D. Perera and P. Harrowell, Phys. Rev. E 59, 5721 (1999).

12. H. Fynewever, D. Perera and P. Harrowell, J. Phys: Cond. Mat. 12, A399 (2000).

13. F.G. Padilla and P. Harrowell, J. Chem. Phys. 116, 4232 (2002).

14. N. Fisher, T.Lewis and B. Embleton, Statistical Analysis of Spherical Data. (Cambridge University Press, Cambridge, 1987). We have used the implementation of this interpolation in the Origin 7.5 plotting software.

15. G. K. Kanji, 100 Statistical Tests. (SAGE Publications, London, 1999).

16. T. Odagaki, and Y. Hiwatari, Phys. Rev. A 43, 1103 (1991).

17. M. Hurley and P. Harrowell, J. Chem. Phys. 105, 10521 (1996).

18. L. Buisson, L. Bellon and S. Ciliberto, J. Phys. Cond. Matter 15, S1163 (2003).

19. A. Crisanti and F. Ritort, Europhys. Lett. 66, 253 (2004).

20. L. Cipelletti and L. Ramos, J. Phys. Cond. Matter 17, R253 (2005).

21. X.H. Qiu amd M.D. Ediger, J. Phys. Chem. B 107, 459 (2003).

22. A.Widmer-Cooper and P.Harrowell, unpublished results.

**Figure Captions**

Figure 1. The distribution of total energy per particle for the 10 initial configurations at T=0.4 in the soft disk mixture with their original momenta (dashed line) and for 5 initial configurations, each with 100 random assignments of momenta from the Maxwell-Boltzmann distribution.

Figure 2. The mean, range, standard deviation (stdev) and the ratio stdev/mean for the propensity distributions calculated for ten configurations each at T = 0.4, 0.46, 0.5, 0.6, 0.8, and 1. At each temperature, the configurations were separated by $75\tau_e$ from each other, and the propensities were averaged over 100 runs of $1.5\tau_e$. Note the different *y*-axis scales.

Figure 3. The propensity distributions over small and large particles for selected configurations at (a) *T* = 1 and (b) *T* = 0.4. The propensities were averaged over 100 runs of $1.5\tau_e$.

Figure 4. The spatial distribution of propensities at *T* = 0.4 for four configurations separated by $75\tau_e$. The propensities were averaged over 100 runs, and the scale is the same as in Figure 7b.

Figure 5. Cluster measures of spatial heterogeneity for particles with propensities in the top 10%. Data points are shown individually for ten configurations each at *T* =

0.4, 0.5, 0.6, 0.8 and 1. Statistics obtained using random values are shown for comparison. The dotted line represents the maximum variance possible for a given number of clusters (see text for more details).

Figure 6. Convergence of the relative uncertainty in the propensity $R$ (see Eq. 2) as a function of the total number of runs $N_{runs}$ for configurations at (a) $T = 0.4$ and (b) $T = 1$. The error bars indicate the range of $R$ values at a given number of runs, and the curve joins the mean values of $R$, where the average is taken over particles.

Figure 7. Convergence of the spatial distribution of propensity as a function of the number of runs for a configuration at $T = 0.4$. The propensities were calculated using (a) 50 runs and (b) 1000 runs. Note that there is little difference in the coarse grained spatial variation between the two plots.

Figure 8. The propensities and their 95% confidence intervals for particles along a line parallel to the $x$-axis in a configuration at $T = 0.4$. The propensities and their uncertanties were calculated using 1000 runs. Note that the error bars are just under twice the standard error.

Figure 9. The distribution of particle displacements, $f_i(\Delta r)$, over an isoconfigurational ensemble of 100 runs for a single particle at $T = 0.4$. Note the highly asymmetric and non-Gaussian shape of the distribution.

Figure 10. The distribution of single particle non-Gaussian parameters $\alpha_i$ (see Eq. 3) for configurations at $T = 0.4$ and 1.0, calculated using ensembles of 1000 runs.

Figure 11. (a) The spatial distribution of the single particle non-Gaussian parameter $\alpha_i$ for a configuration at $T = 0.4$. The white circles indicate the positions of large particles with six large neighbours. (b) The propensity map for the same configuration used in (a). Quantities were calculated using ensembles of 100 runs.

Figure 12. The particle directionality $d_i$ as a function of propensity for ten configurations each at (a) $T = 0.4$ and (b) $T = 1$. Quantities were calculated using 100 runs.

Figure 13. Displacement vectors for selected particles at $T = 0.4$ with high directionality and either (a) low or (b) high propensity. The vectors are from iso-configurational ensembles of 100 runs.

Figure 14. The correlation between the motion of a particle $i$ and all other particles $j$ as a function of the distance between $i$ and $j$. The moving average has been indicated by a thick line, and the Pearson's correlation coefficient $K$ between displacement magnitudes was calculated using data from an ensemble of 100 runs.

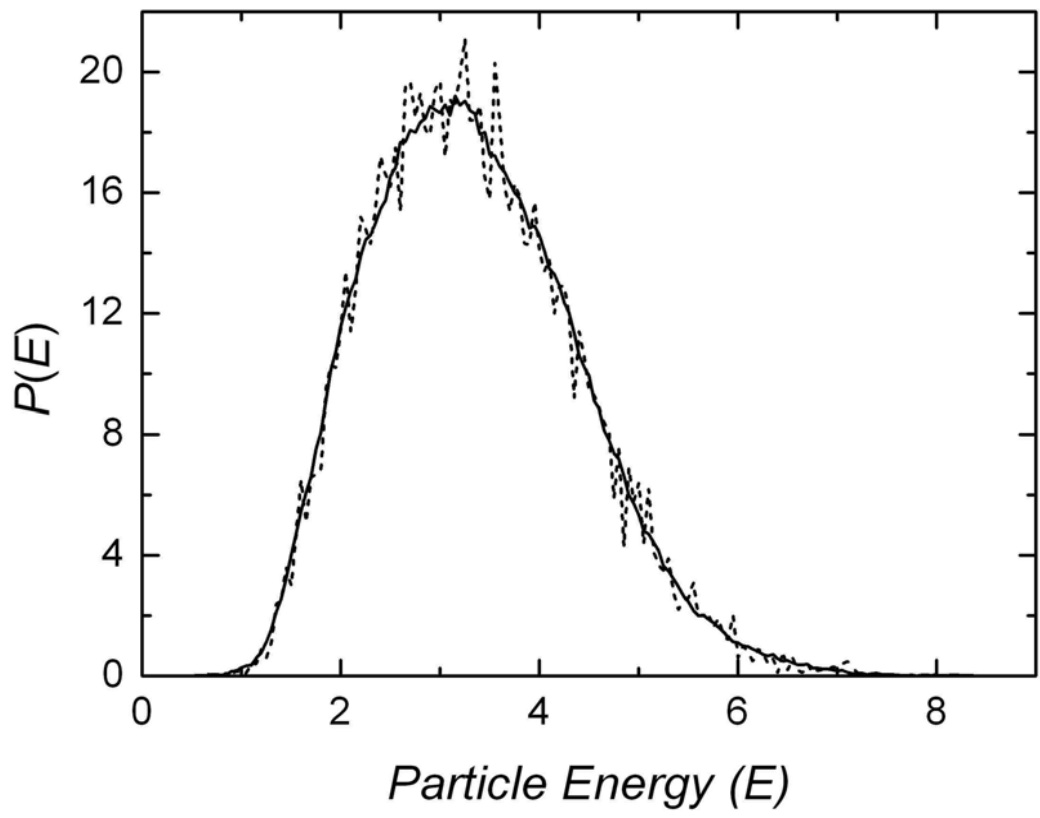
Figure 1

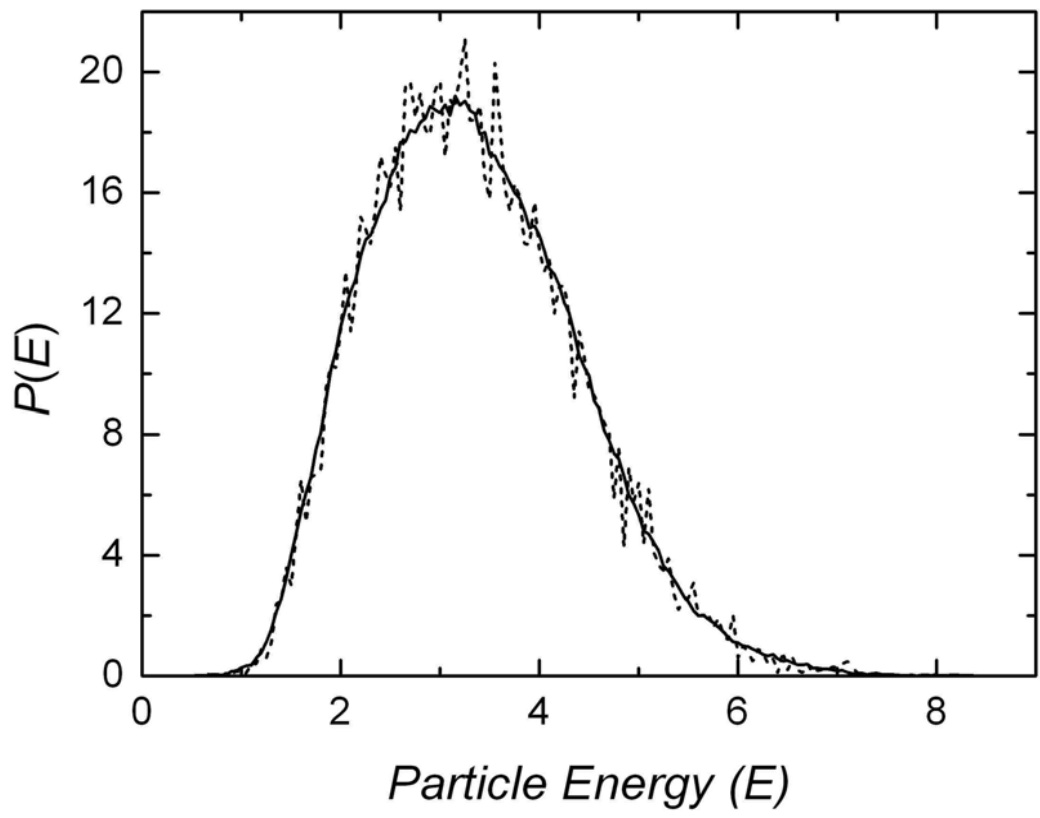

Figure 1

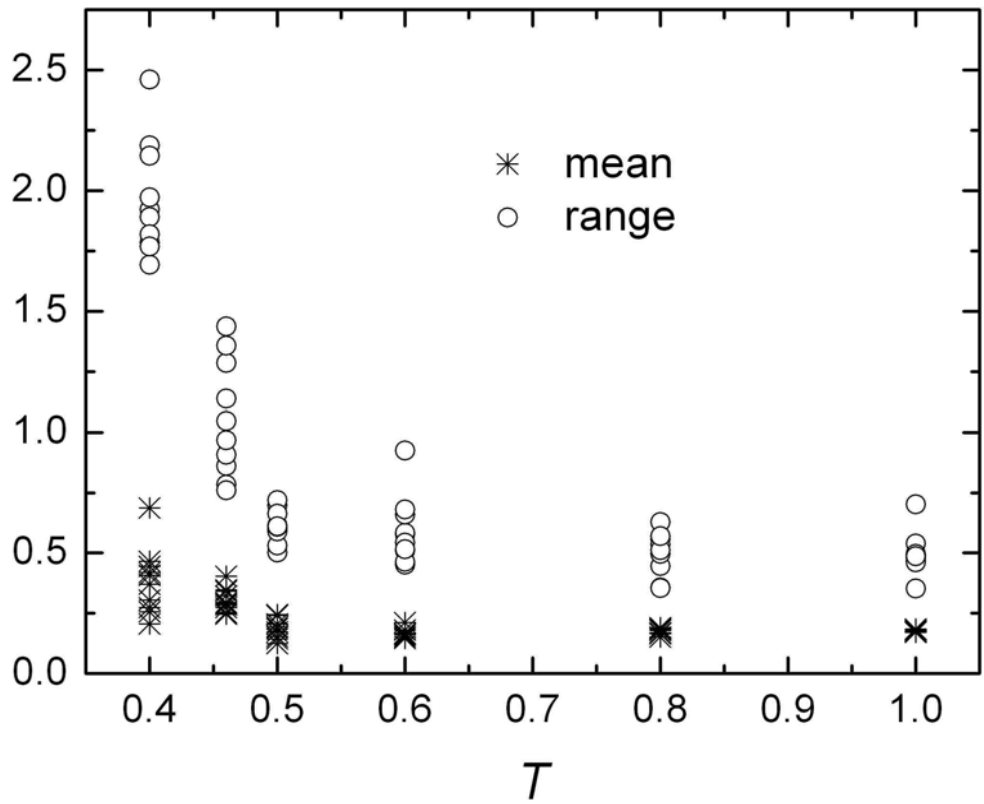

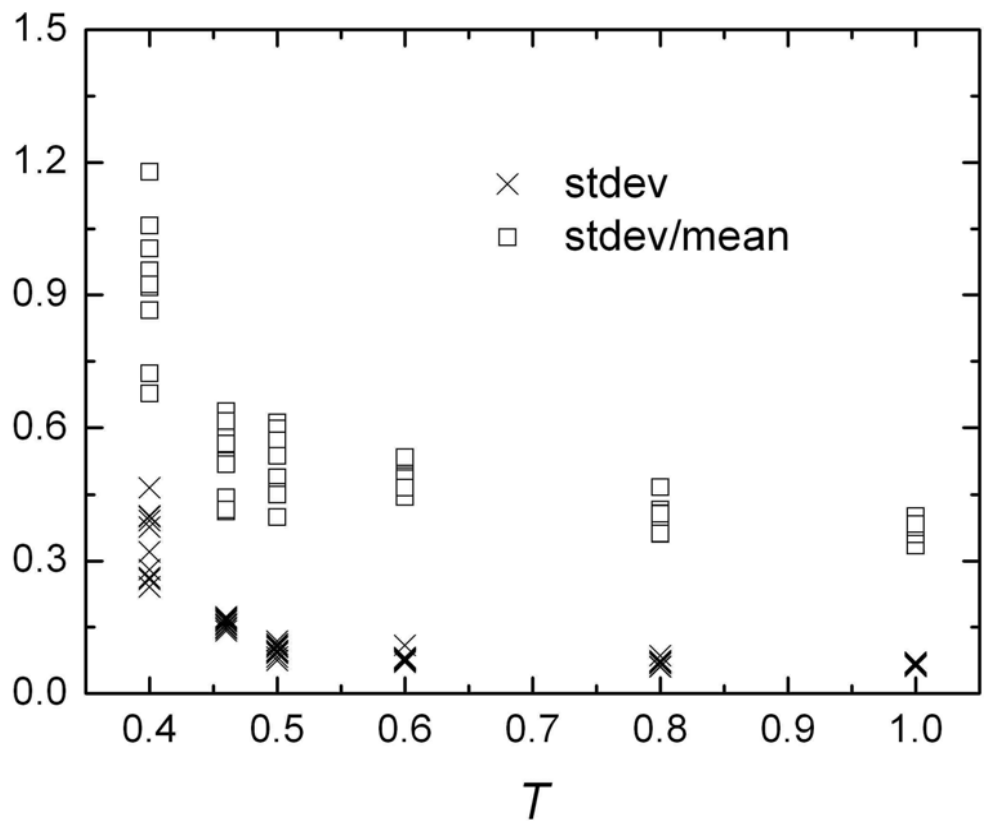

Figure 2.

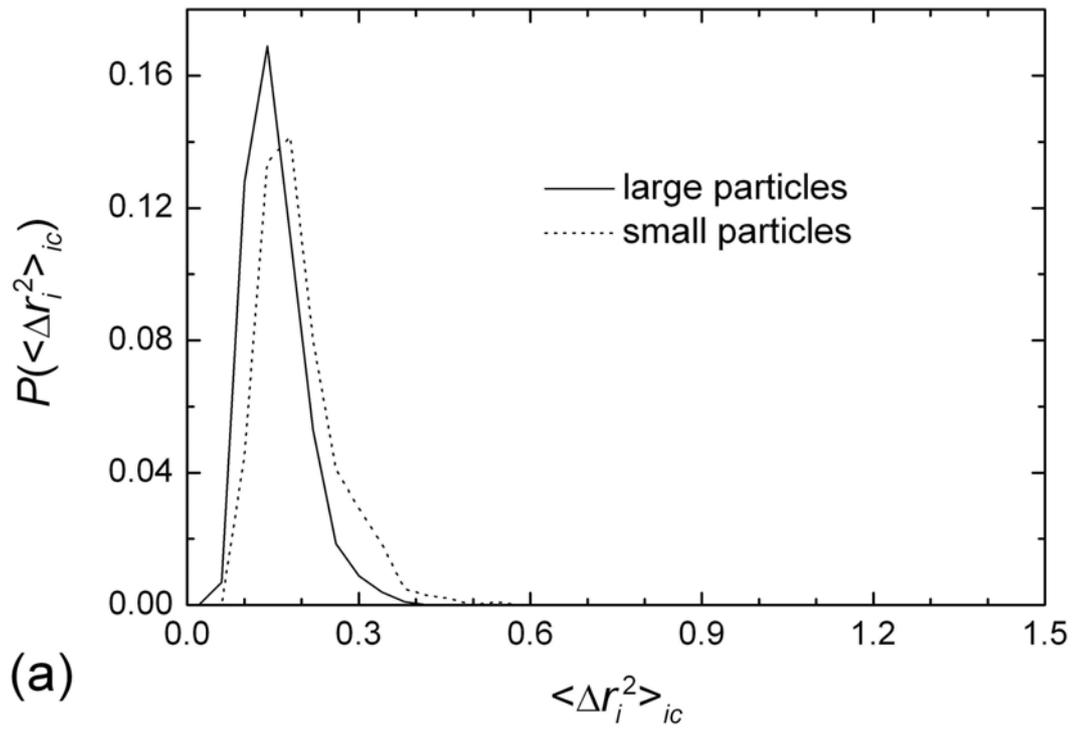

(a)

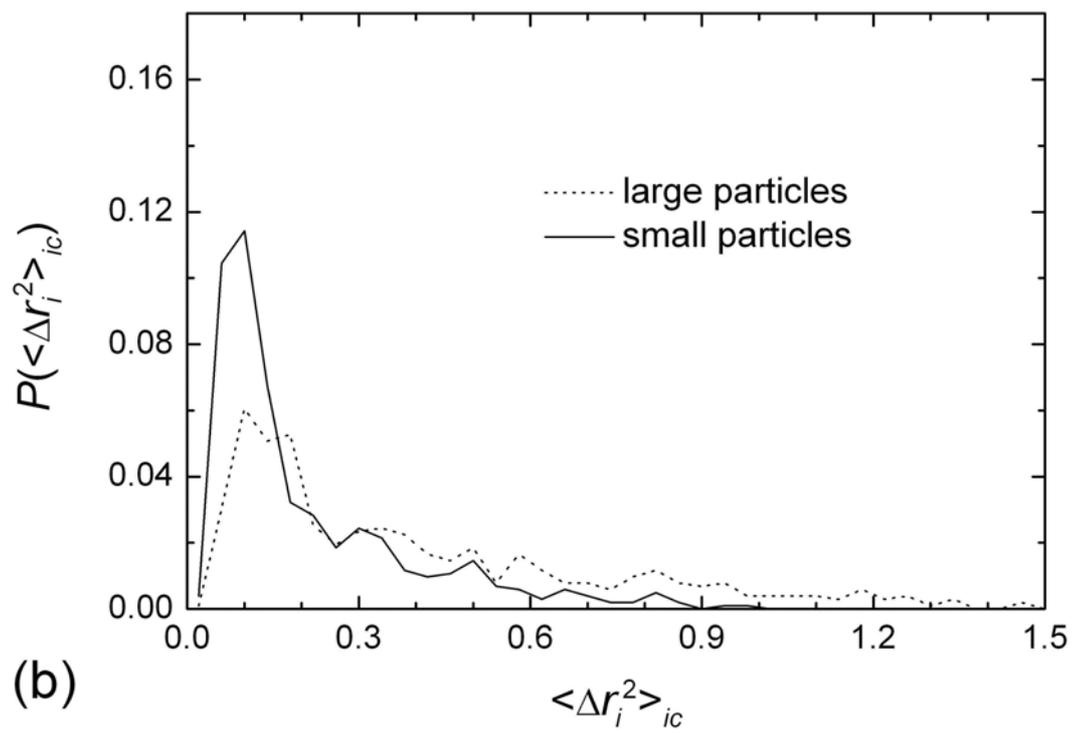

(b)

Figure 3.

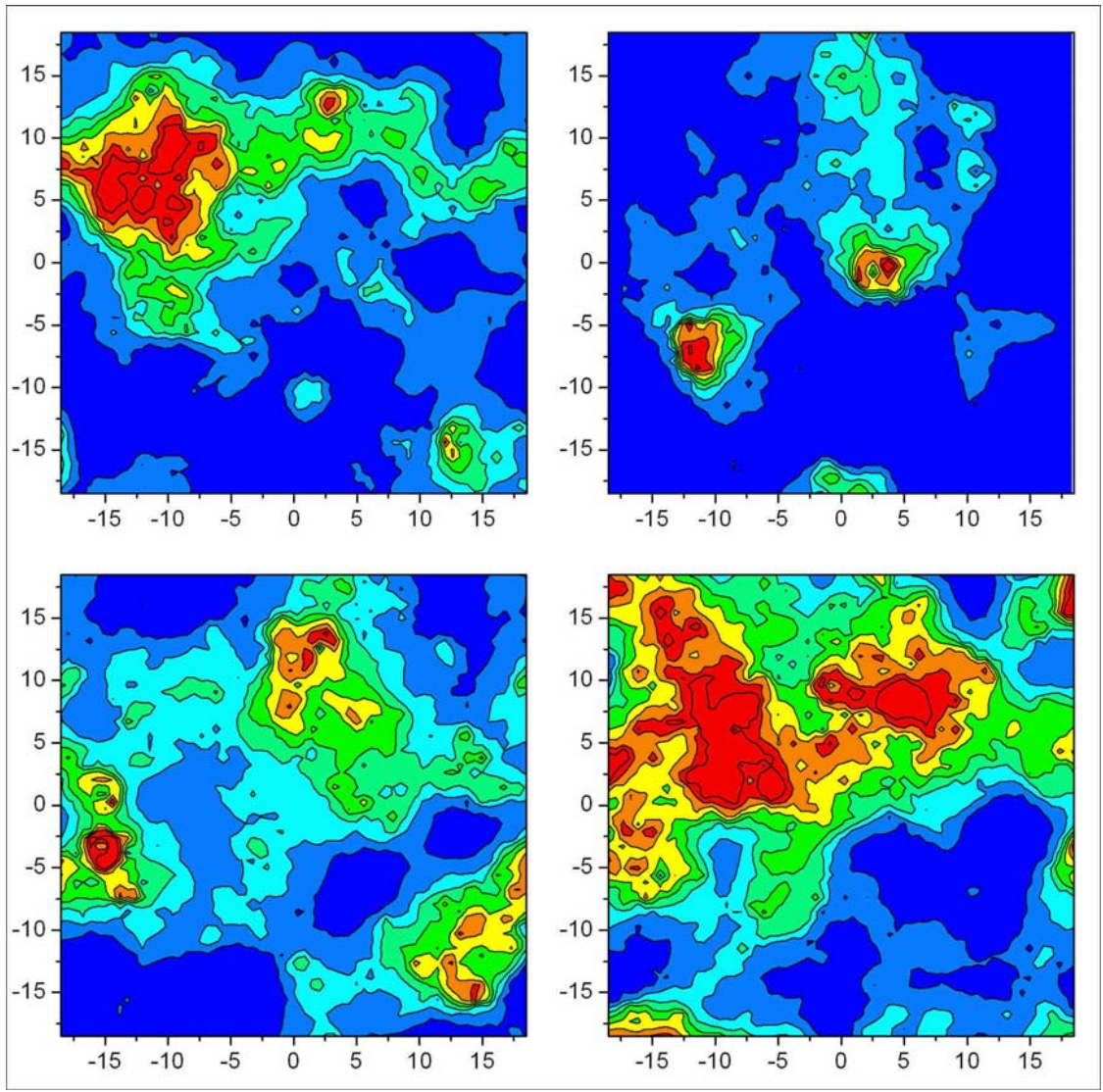

Figure 4.

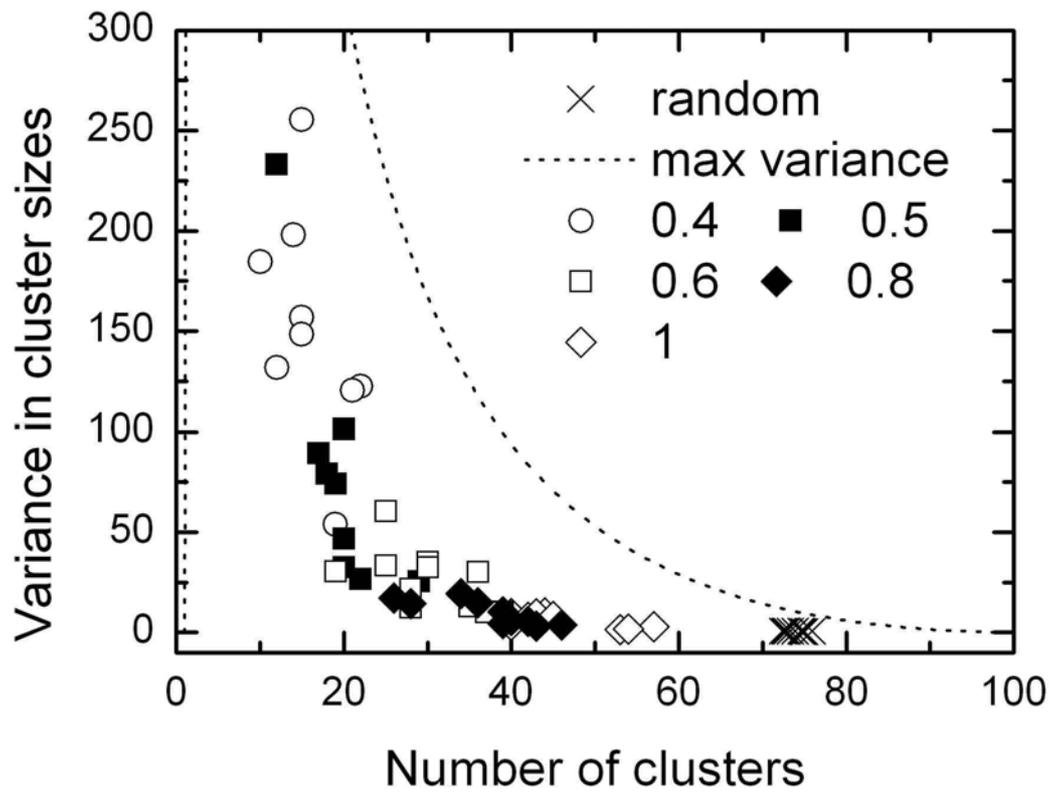

Figure 5.

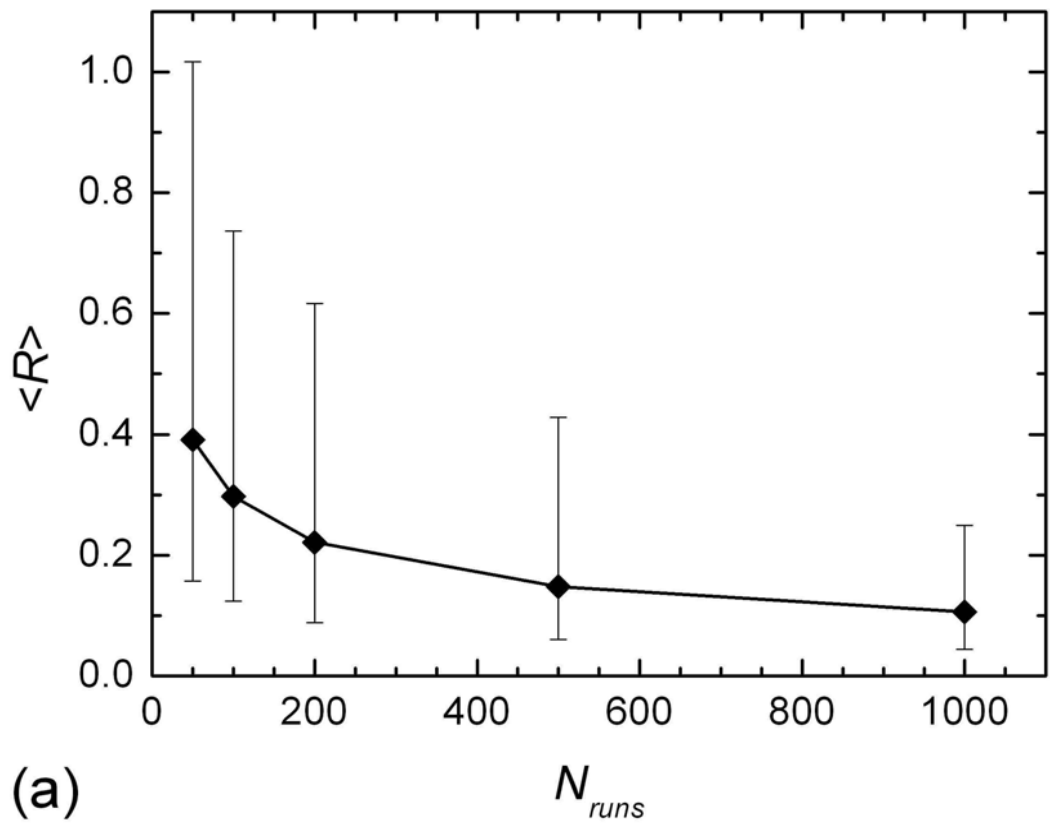

(a)

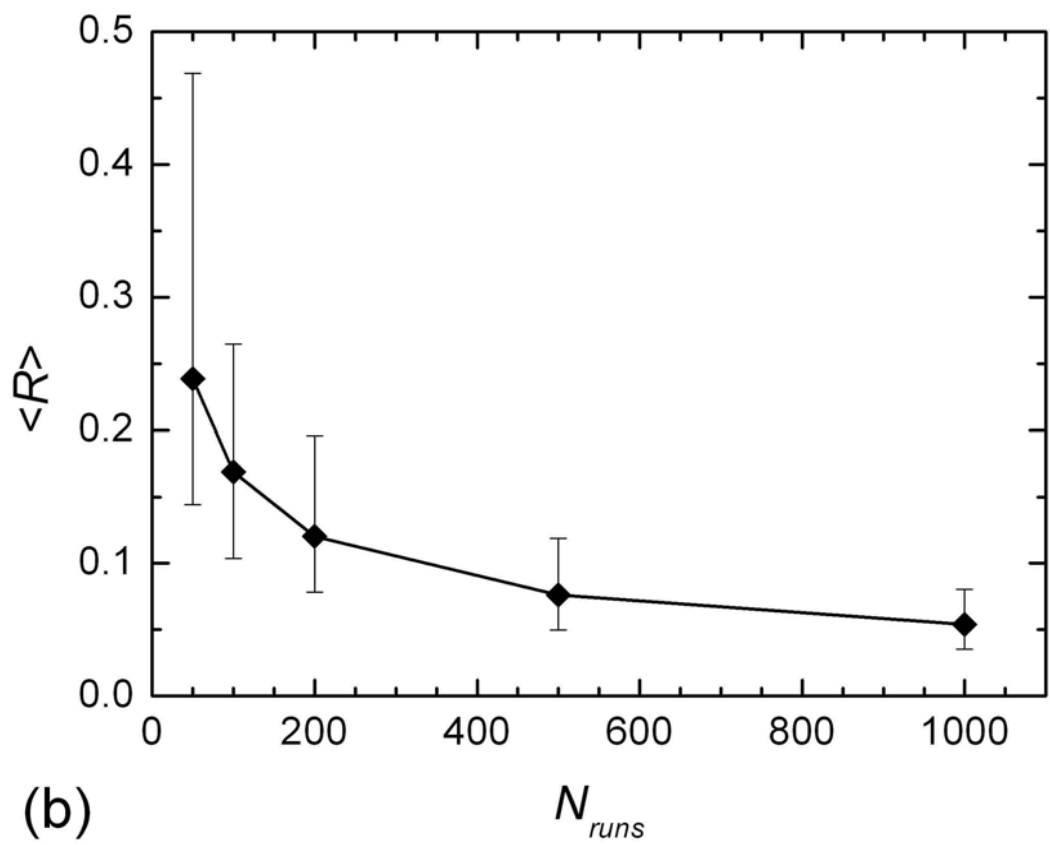

(b)

Figure 6.

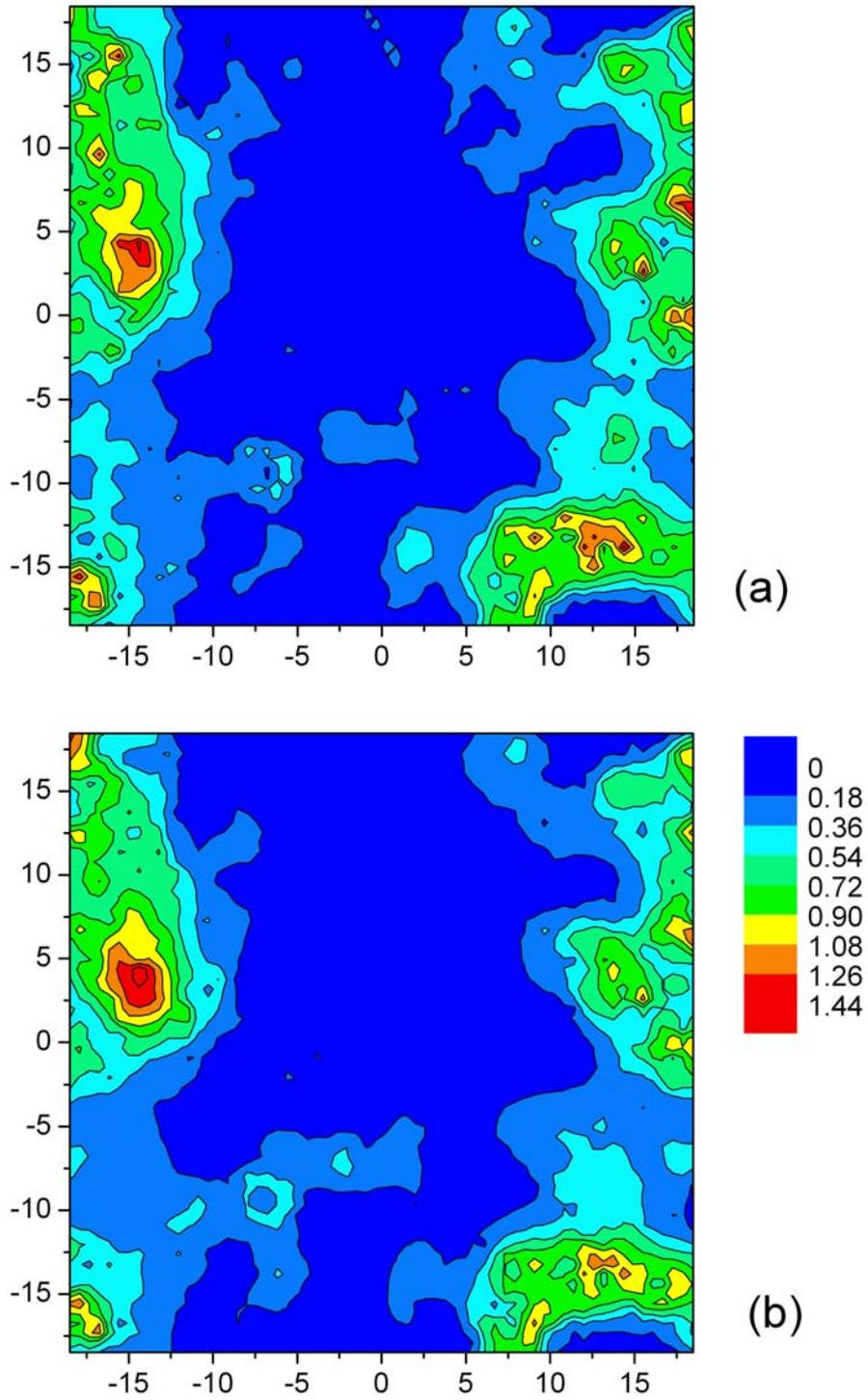

Figure 7.

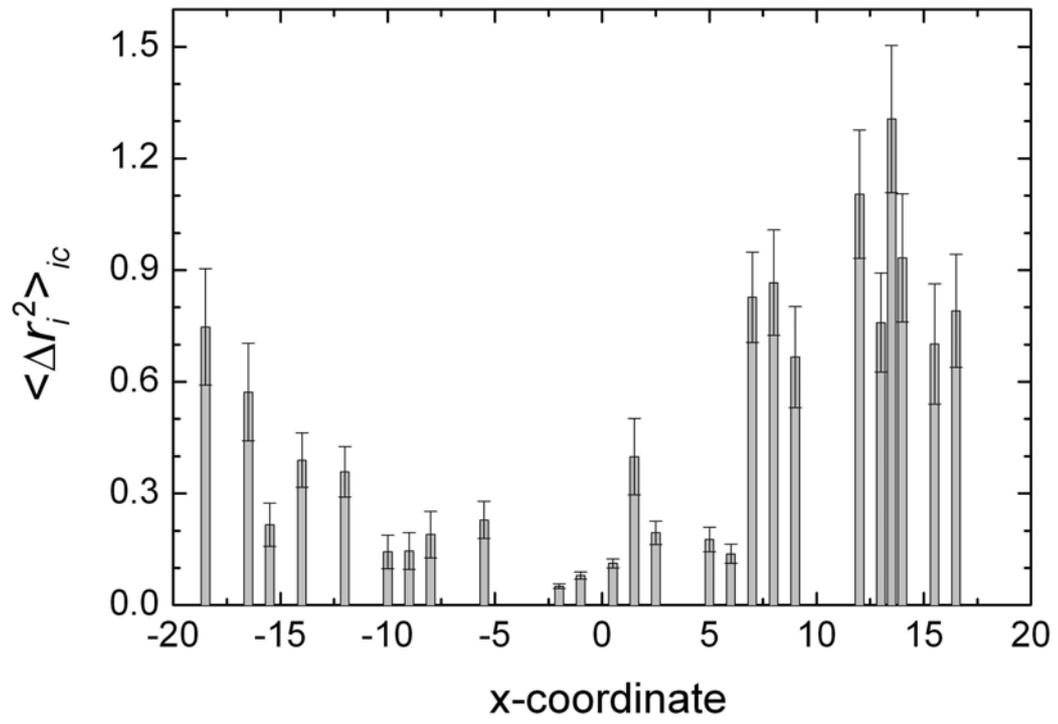

Figure 8.

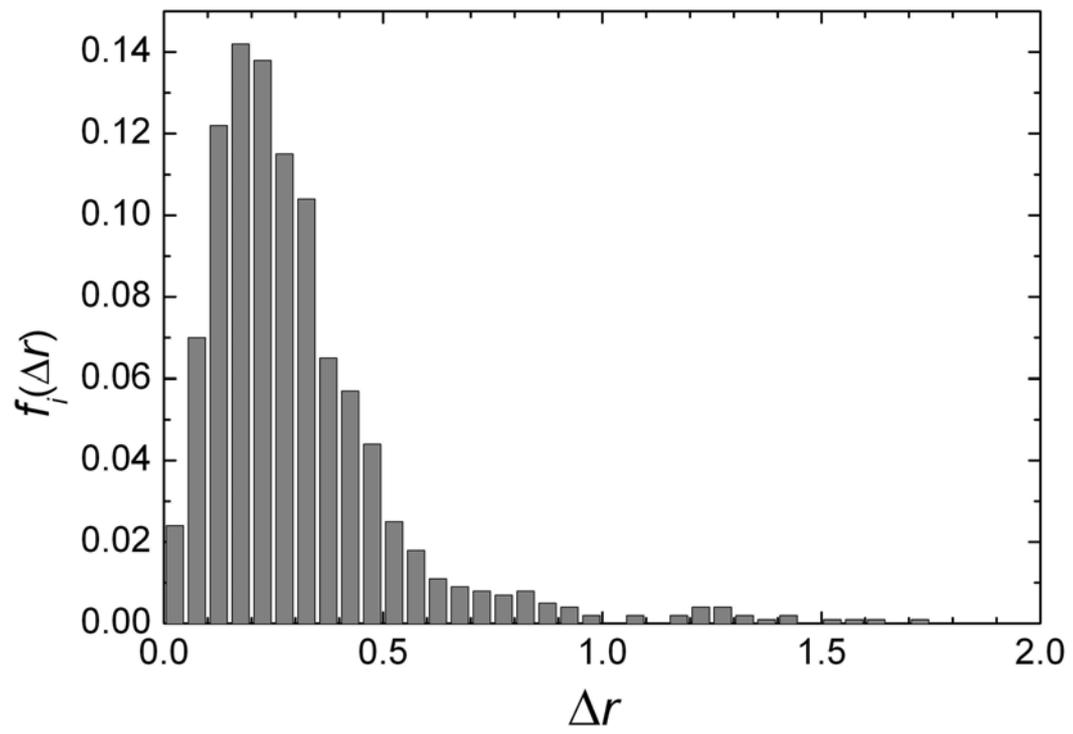

Figure 9.

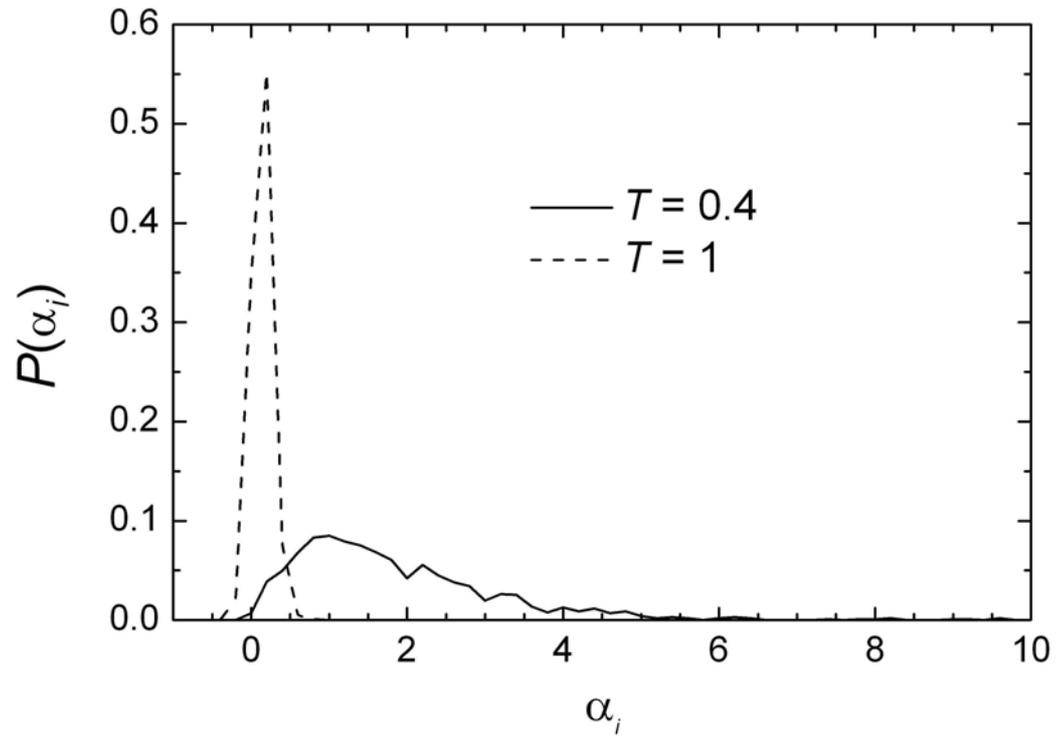

Figure 10.

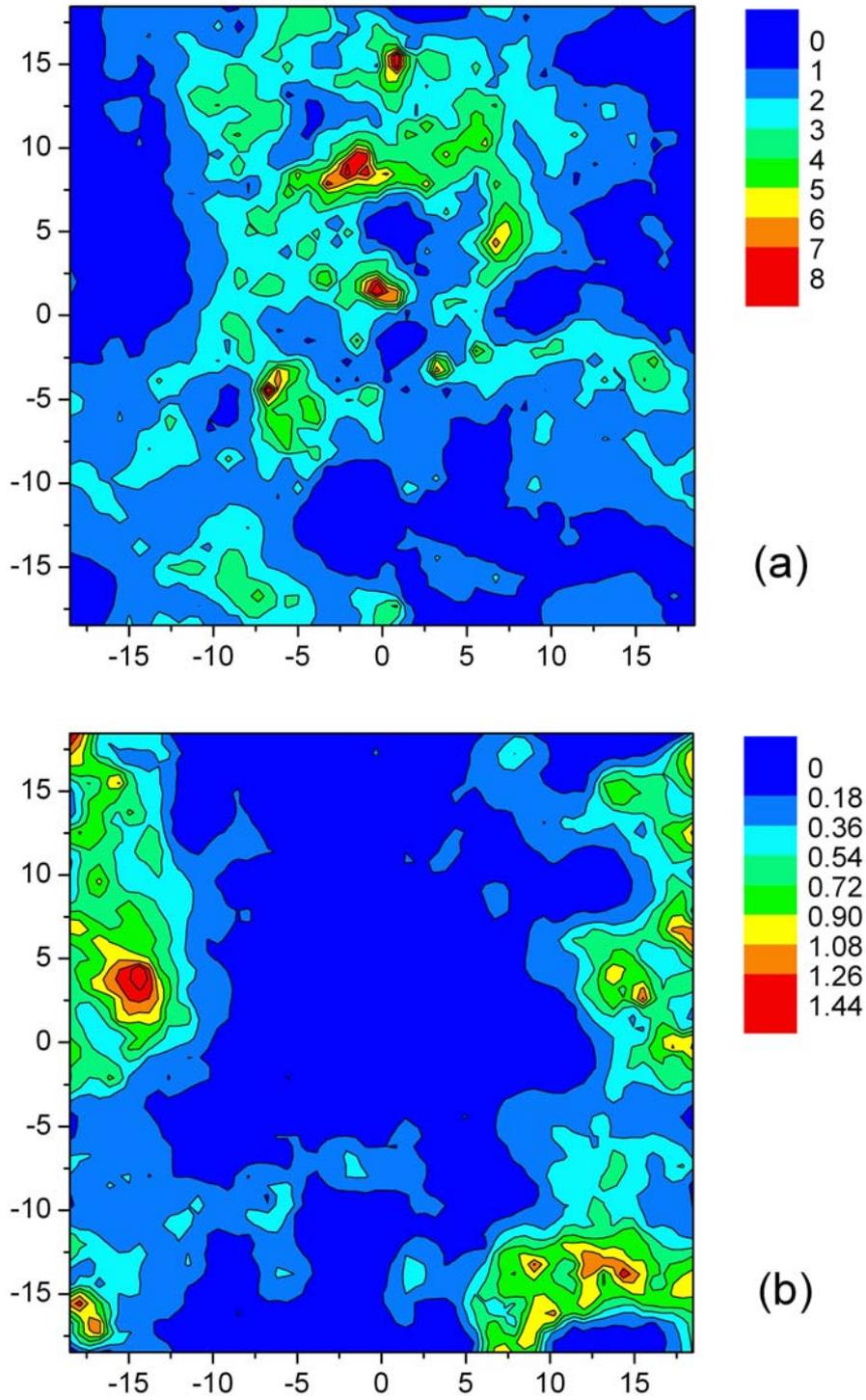

Figure 11.

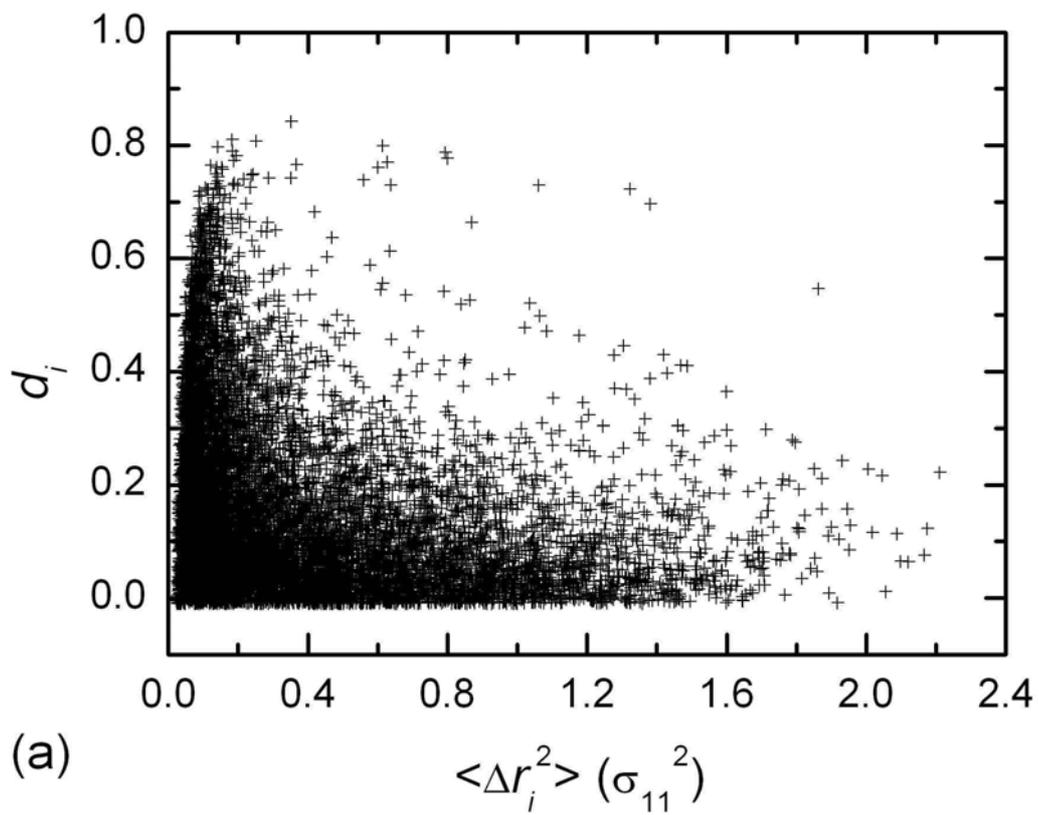

(a)

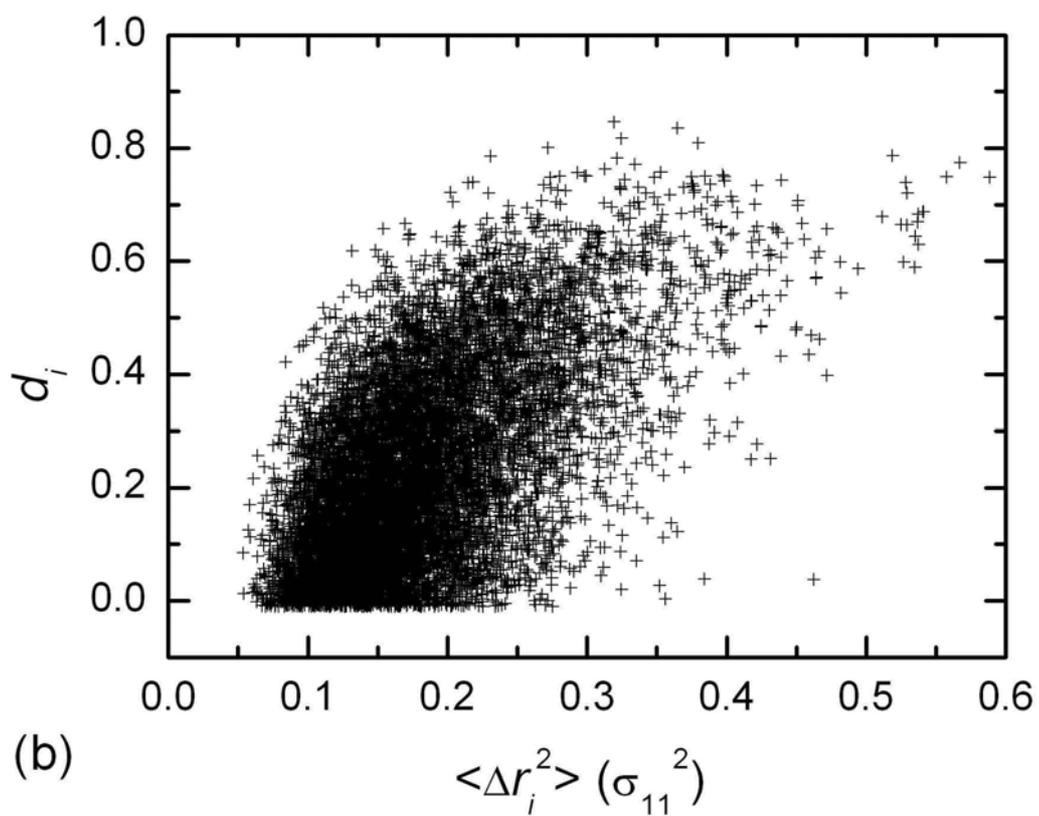

(b)

Figure 12.

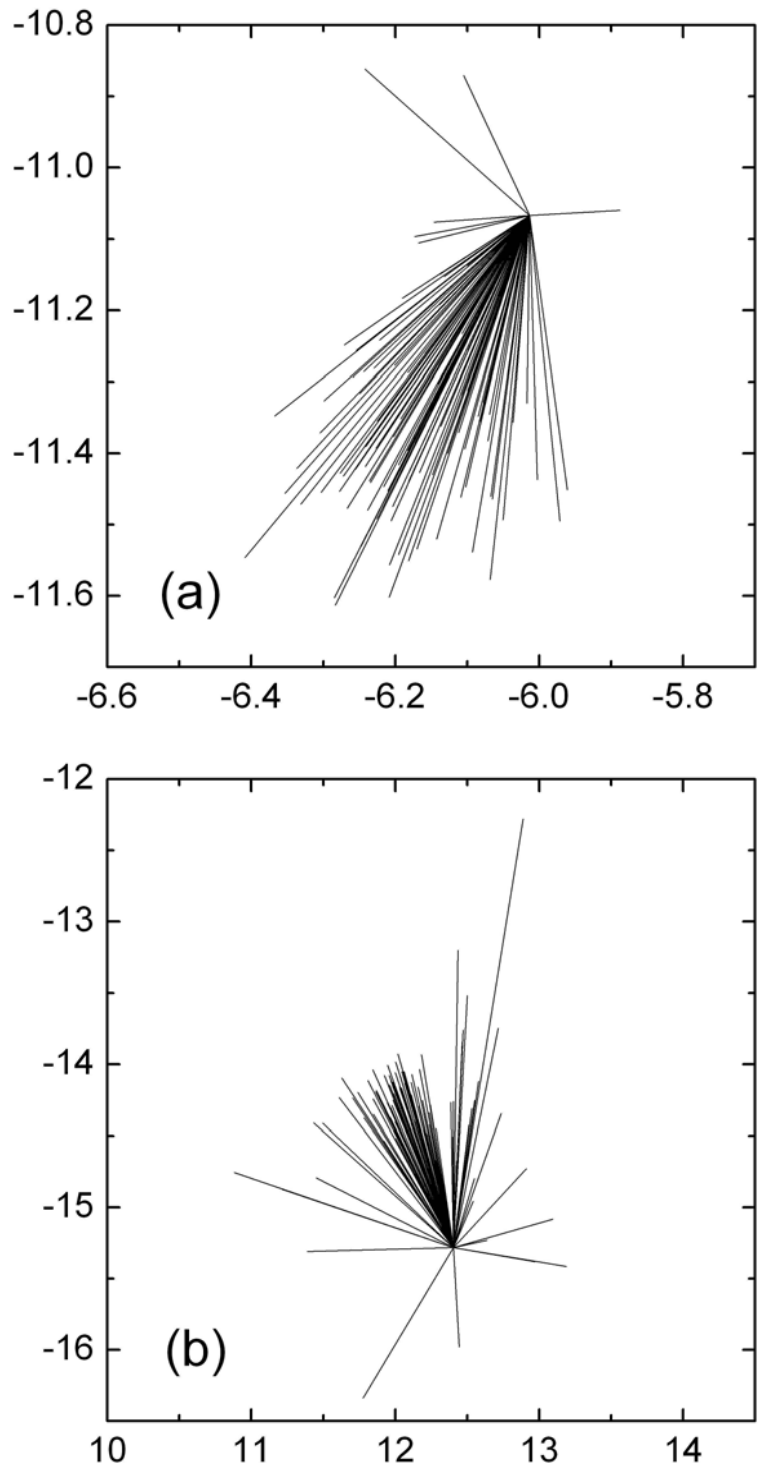

Figure 13.

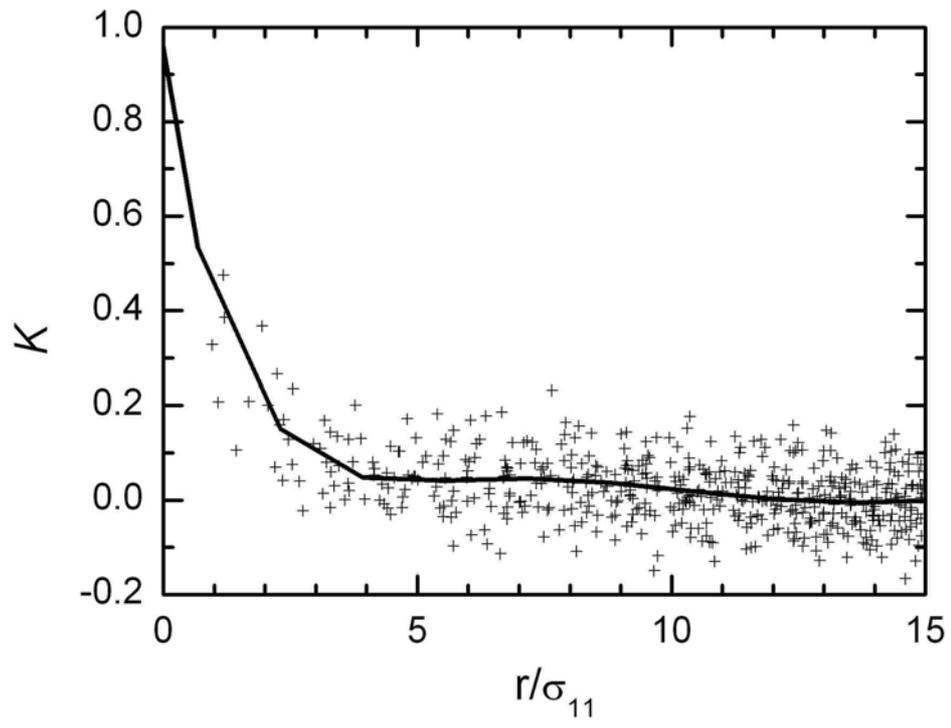

Figure 14.